\documentclass[twocolumn]{aastex62}
\usepackage[version=4]{mhchem}
\usepackage{booktabs} % Allows the use of \toprule, \midrule and \bottomrule in tables for horizontal lines
\usepackage{graphics, multirow}
\usepackage{enumitem}
\usepackage{refcount}
\newcommand{\cmmnt}[1]{}

\graphicspath{ {./figures/} }

\accepted{December 21, 2018}
\submitjournal{The Astrophysical Journal}

\shorttitle{The Influence of \ce{H2O} Pressure-Broadening ...}
\shortauthors{Gharib-Nezhad $\&$ Line}
\usepackage{subfigure}
\begin{document}

\title{The Influence of \ce{H2O} Pressure Broadening in High Metallicity Exoplanet Atmospheres}

%% The \author command is the same as before except it now takes an optional
%% arguement which is the 16 digit ORCID. The syntax is:
%% \author[xxxx-xxxx-xxxx-xxxx]{Author Name}
%%
%% This will hyperlink the author name to the author's ORCID page. Note that
%% during compilation, LaTeX will do some limited checking of the format of
%% the ID to make sure it is valid.
%%

%\correspondingauthor{????}
%\email{???}

\author{Ehsan Gharib-Nezhad}
\affil{School of Molecular Sciences, Arizona State University, Tempe, AZ 85287, USA.}

\author{Michael R. Line}
\affiliation{School of Earth and Space Exploration, Arizona State University, Tempe, AZ 85287, USA.}

\begin{abstract}
Planet formation models suggest broad compositional diversity in the sub-Neptune/super-Earth regime, with a high likelihood for large atmospheric metal content ($\geq$ 100 $\times$ Solar). With this comes the prevalence of numerous plausible bulk atmospheric constituents including \ce{N2}, \ce{CO2}, \ce{H2O}, \ce{CO}, and \ce{CH4}. Given this compositional diversity there is a critical need to investigate the influence of the background gas on the broadening of the molecular absorption cross-sections and the subsequent influence on observed spectra.  This broadening can become significant and the common \ce{H2}/He or ``air''  broadening assumptions are no longer appropriate.  In this work we investigate the role of water self-broadening on the emission and transmission spectra as well as on the vertical energy balance in representative sub-Neptune/super-Earth atmospheres.  We find that the choice of the broadener species can result in a 10 -- 100 parts-per-million difference in the observed transmission and emission spectra and can significantly alter the 1-dimensional vertical temperature structure of the atmosphere. Choosing the correct background broadener is critical to the proper modeling and interpretation of transit spectra observations in high metallicity regimes, especially in the era of higher precision telescopes such as JWST. 

\end{abstract}


\section{Introduction}

%% Planetary formation
A primary goal of exoplanet science is the determination of basic planetary conditions.  Transit spectrophotometry observations of planetary atmospheres offer a window into fundamental quantities such as climate and composition e.g \cite{Madhusudhan2016}).  Determining atmospheric composition is a necessary requirement for assessing the relative importance of various chemical processes \citep{Moses2014} and greatly assists in understanding planet formation by linking volatile inventory to proto-planetary disk processes \citep{Oberg2011, Mordasini2016}.

One of the key findings of the Kepler Mission \citep{Borucki1997} is that a majority of exoplanets fall within this ``warm sub-Neptune'' regime ($\sim$2\textbf{\textendash}4 Earth radii, T$<$1000 K ) \citep{Fressin2013}. These planets have been an intense area of focus for transit spectra observations with the Hubble Space Telescope (HST) \citep{Kreidberg2014a,Fraine2014,Knutson2014a} and will be over the next decade  as they serve as the link between jovian worlds and terrestrial planets as well as being the most prolific population of planets to be found by the Transiting Exoplanet Explorer Satellite (TESS, \citep{Sullivan2015,Louie2018,Barclay2018,Kempton2018}). 

Planet formation, interior structure, and atmospheric chemistry modeling \citep{Fortney2013,Moses2013,Lopez2014} suggest extreme compositional diversity within this sub-population, with a high likelihood for large atmospheric metallicities (>300$\times$ Solar).  Given this potential for compositional diversity, the assumption of ``jovian-like'' \ce{H2}/He-dominated atmospheres may not always be appropriate.  Instead, with currently measured atmospheric metallicities reaching as high as $\sim$300-1000$\times$ solar \citep{Line2014, Fraine2014, Kreidberg2014, Knutson2014a, Morley2017}, molecules such as \ce{H2O} and \ce{CO2} will become the dominant bulk constituents \citep{Moses2013,Hu2014}. 

Along with this diversity in composition, comes with it numerous challenges in atmospheric modeling ranging from chemical modeling \citep{Hu2014} to cloud microphysics \citep{Ohno2018} to 3D climate modeling \citep{Kataria2014}. Nearly all flavors of atmospheric modeling that aim to make observational predictions necessarily require radiative transfer computations.  A key necessary ingredient in radiative transfer computations is the opacities, which for planets, are dominated by the molecular absorption cross-sections (hereafter, ACS). The ACS of a given molecule typically consist of billions of lines representing the ability of a molecule to absorb or emit photons. Each line has its own line-width (or broadening) typically specified through the degree of thermal/Doppler and pressure broadening \citep{Goody1995}. Pressure broadening is the net cumulative effect of interactions between the absorbing molecule in question  (e.g.,  \ce{H2O}) with its neighboring molecules (or bath gases, e.g., \ce{H2}, He) or by self-broadening (\ce{H2O} with itself). Much exo-atmospheric relevant ACS focus, specifically broadening, has been jovian-centric (e.g., \ce{H2}$/$\ce{He} dominated compositions and broadening \cite{Freedman2008,Tennyson2016,Grimm2015,Hedges2016}) being largely driven by the abundance of high fidelity ``hot-Jupiter'' observations and carry over from brown dwarf modeling. 

Exploration of pressure broadening assumptions in exo-atmospheres is not new (e.g., \cite{Grimm2015,Hedges2016}).  \cite{Hedges2016} provide a comprehensive overview of the various pressure broadening effects including resolution, line-wing cutoff, Doppler vs. pressure, and more relevant to our investigation, an initial look at the impact of a broadener choice. They too explore the impact of \ce{H2O} vs \ce{H2} broadening on the \ce{H2O} ACS, specifically over HST wavelengths, and find that the band-averaged ACS can change by up to an order-of-magnitude.   

In this letter we expand upon the work in \cite{Hedges2016} to not only determine the influence of \ce{H2O} self-broadening on the \ce{H2O} ACS, but also as a function of water fraction, and more importantly we quantitatively assess the integrated effect that the broadener choice has on the {\it observable spectra} as well as on the impact on atmospheric vertical energy balance.  This work is crucial to the proper interpretation of transit spectra observations in high metallicity regimes, expected of the sub-Neptune/Super-Earth population.   In \S \ref{Methods} we describe our data sources and how we compute the ACS and the transmission/emission spectrum and self-consistent modeling approach. In \S \ref{Results} we compare the impact of \ce{H2O} self-broadening with the standard \ce{H2}/\ce{He} broadening assumption.   Finally, in \S \ref{Conclusions} we discuss the implications and future prospects. We also make our newly computed water ACS grid for both broadeners publically available\footnote{LINK:TBD UPON ACCEPTANCE}.

\section{Methods}\label{Methods}
%% The lack of Pressure-broadened ACS
In this initial investigation on the impact of {\it non} \ce{H2}/\ce{He} foreign broadening on transmission/emission spectra, we choose to focus on \ce{H2O} because: 1) \ce{H2O} is the most prominent absorber in exoplanet spectra due to its large abundance over a range of elemental compositions 
\citep{Moses2013} and multiple strong absorption bands from the optical to far infrared wavelengths and 2) it shows the largest sensitivity to choice of broadener when compared to other species (a factor of $\sim$7 increase in broadening when compared to \ce{H2/He}, Table \ref{tab: Table1}).

%TABEL 1******************************************
\begin{table}[h!tb]
  \footnotesize % text size of table content
  \caption{Lorentzian half-width coefficients $\gamma_L$ [cm$^{-1}$/bar]$^{*,\dagger}$ for relevant broadeners. The focus of this work is on influence H$_2$O self and H$_2$/He broadening on the H$_2$O absorption cross-sections (bold). }\label{tab: Table1}   
   \centering % center the table
\begin{tabular}{ c|c|c|c } 
\hline
Absorber & Broadener & $\gamma_L$   &  $relative$\ $to$\ $ \gamma_L^{H_2/He}$ \\
\hline\hline
\multirow{3}{4em}{\ce{H2O}} 
& \bf{Self} $^\ddagger$	&  \bf{$\ $0.3 -- 0.54} 	& \bf{7$\times$} \\ 
& \bf{\ce{H2}/\ce{He}} $^\mathparagraph$  	& \bf{0.05 -- 0.08} 	& \bf{1$\times$}  \\ 
& \ce{CO2} 			& 0.15 -- 0.20 	& 3$\times$  \\ 
%& \ce{CO} 			& ??? 			& $\times$ 	\\
%& \ce{CH4} 			& 0.08 			& $\times$  \\
& \ce{air} 			& 0.08 -- 0.1$\ $ 	& 1.5$\times$  \\ 
\hline

\multirow{3}{4em}{\ce{CH4}} 
&    Self  			& 0.06 -- 0.09 	& 1.5$\times$ 	\\ 
& \ce{H2}/\ce{He} 	& 0.05 -- 0.08 	& 1$\times$	\\ 
& \ce{H2O} 			& 0.06 -- 0.09 	& 1.5$\times$  	\\ 
& \ce{CO2} 			& 0.07 -- 0.09 	& 1.5$\times$  	\\ 
&    air   			& 0.02 -- 0.07 	& 1$\times$ 	\\ 
\hline

\multirow{3}{4em}{\ce{CO2}} 
& Self    			& 0.08 -- 0.12  	& 2$\times$ \\ 
& \ce{H2}/\ce{He} 	& 0.09 -- 0.12   & 2$\times$   	\\ 
& \ce{H2O}			& 0.10 -- 0.14  	& 2.5$\times$  	\\ 
& air     			& 0.05 -- 0.08 	& 1$\times$ 	\\ 
\hline

\multirow{3}{4em}{\ce{CO}} 
& Self    			& 0.04 -- 0.09  	& 1$\times$ 	\\ 
& \ce{H2}/\ce{He} 	& 0.04 -- 0.08	& 1$\times$  	\\ 
& \ce{H2O}			& 0.07 -- 0.1$\ $ 	&  1.5$\times$ 	\\ 
& \ce{CO2}			& 0.09 -- 0.1$\ $ 	&  1.5$\times$ 	\\ 
& air     			& 0.05 -- 0.07  	& 1$\times$ 	\\ 
\hline\hline
\end{tabular}
       \scriptsize
       \hfill\parbox[t]{\linewidth}{$^*$ The pressure broadening/Lorentzian line profile is defined with a half-width $\Gamma_L$ = $(T/296)$ $^{-n_T}$   $\sum_{b}$  $\gamma_L^b$ $P_{b}$ 
                             where $\gamma_L^b$ is the Lorentzian coefficient for broadener, $b$, $P_{b}$ is the partial pressure of broadener $b$, and n$_{T}$ is the thermal coefficient (typically 0.5 under 								 kinetic theory).\\
                             $^\dagger$ Data extracted from Refs. in Table 3 of \citep{Hartmann2018} and from \citep{Gordon2017}.\\
                             $^\ddagger$ Denoted by \ce{H2O}@[self] in the text and figures.\\
                             $^\mathparagraph$ Denoted by \ce{H2O}@[\ce{H2}+\ce{He}] in the text and figures.}

\end{table}    
%************************************************

The approach is to compute the \ce{H2O} ACS under different end-member compositional scenarios, with the first the standard ``jovian-like'' \ce{H2}/\ce{He} broadening (\ce{H2O}@[\ce{H2}+\ce{He}]) and the second, pure \ce{H2O} broadening (\ce{H2O}@[self]), which is more appropriate for high metallicity or all-steam atmospheres.  We then determine the spectral differences between \ce{H2}/\ce{He} and self-broadening of \ce{H2O} in representative atmospheres. 

%\subsection{Line Lists}
%AAAA           In order to generate our \ce{H2O}@[\ce{H2}+\ce{He}] and \ce{H2O}@[self]  ACS database, we utilize the freely available EXOMOL \citep{Tennyson2016} line-list data which provides the full BT2 line list \citep{Barber2006}; valid over a broad range of ``typical" exoplanet environmental conditions. 

%%%%%%%%%%%%%%%%%%%%%%%%%%%%%%%%%%%%%%%%%%%%%%%%%%%%%%%%%%%%%%%%%%%%%%%%%%%%%%%%%%%%%%%%%%%%%%%%%%%%%%%%%%%%%%%%%%%%%%%%%%%%%%%%%%%%%%%%%%%%
\subsection{Computation of pressure-broadened \ce{H_2 ^{16}O} absorption cross-section} 
%TABEL 2******************************************
\begin{table}[ht]
  \footnotesize % text size of table content
  \caption{ Grid and computational assumptions over which the H$_2 ^{16}$O cross sections are computed.  There are 270 T-P combinations and two broadener choices (H$_2$+He vs. H$_2$O).  A variable wavenumber resolution is chosen to properly sample the Voigt-widths at each given T-P pair.  Finer sampling results in negligible differences in the ACS. }\label{tab: Table2}     
%   \label{tab:example_multirow}
   \centering % center the table
     \begin{tabular}{p{22pt}p{19pt}p{22pt}p{18pt}p{22pt}p{20pt}p{10pt}p{11pt}p{11pt}}%{lcccccr} % alignment of each column data
       \toprule[\heavyrulewidth]\toprule[\heavyrulewidth]
     \multirow{2}{*}{ACS}
     &Case &1: &  85$\%$ & \ce{H2}  & 15$\%$  & He \\ 
     &Case &2: & 100$\%$ & \ce{H2O}   \\ 
     \midrule[0.3pt]
     \multirow{2}{*}{T(K)}
     &400&425 & 475 & 500  & 575  & 650 &725 &800 \\ 
     & 900 & 1000 & 1100 & 1200& 1300 &1400 & 1500\\
	\midrule[0.3pt]
     \multirow{2}{*}{P(bar)}
 & 10$^{-6}$ & 3$\times$10$^{-6}$ & 10$^{-5}$ & 3$\times$10$^{-5}$  & 10$^{-4}$  & 3$\times$10$^{-4}$ \\ 
 & 10$^{-3}$ & 3$\times$10$^{-3}$ & 10$^{-2}$ & 3$\times$10$^{-2}$  & 10$^{-1}$  & 3$\times$10$^{-1}$ \\ 
 & 1 & 3 & 10 & 30  & 100  & 300 \\ 
    \midrule[0.3pt]
    \multirow{2}{*}{Resolution$^{*}$}
&  &100&  -- & 1000 & cm$^{-1}$ & : &   1/$\Gamma_V$ \\ 
&  &1000&  -- & 30000 & cm$^{-1}$ & : &   2/$\Gamma_V$ \\ 
    \midrule[0.3pt]
     \multirow{2}{*}{Line wing cut-off$^{\dagger}$}
     &&&P$\geq$1&bar:&  100 &cm$^{-1}$\\
     &&&P$<$1& bar:& 300 &cm$^{-1}$\\
     \bottomrule[\heavyrulewidth] 
   \end{tabular}
          \scriptsize
       \hfill\parbox[t]{8cm}{ 
       $^*$ $\Gamma_V$ is the Voigt half-width approximated as 0.5346$\Gamma_L+\sqrt{0.2166 \Gamma_L^2+\Gamma_G^2}$, with $\Gamma_G$ the Doppler width \citep{Olivero1977}.\\  
       $^\dagger$ The Lorentz wing shape may not be appropriate out at such distances \citep{Freedman2008}   }
\end{table}
%BBBB           We utilize the publicly available EXOCROSS\footnote{https://github.com/Trovemaster/exocross} routine \citep{Yurchenko2017} to model the full Voigt profile \citep{Humlicek1979} of every single line over a grid of applicable temperatures and pressures with reasonable sampling resolution and line-wing cutoff (Table \ref{tab: Table2}).  In this study, the pressure-broadened \ce{H2O} ACS  are computed for two set of broadeners: 1)  85$\%$ \ce{H2} and 15$\%$ He using the J-dependent pressure coefficients from EXOMOL \citep{Barton2017a}, and 2) 100$\%$ \ce{H2O} using the average value of available experimental self-broadening coefficients \citep{Ptashnik2016}.  

% COMBINED AAAA &    BBBB !!!!!!!!!
We utilize the freely-available EXOMOL \citep{Tennyson2016} line-list data (e.g., BT2 line list \citep{Barton2017a}) and EXOCROSS\footnote{https://github.com/Trovemaster/exocross} routine \citep{Yurchenko2017} to compute the pressure-broadened \ce{H2O} ACS database (Table \ref{tab: Table2}) for two broadening scenarios: 1)  85$\%$ \ce{H2} and 15$\%$ He (current standard assumption) using the $J$-dependent pressure coefficients from EXOMOL \citep{Barton2017a}, and 2) 100$\%$ \ce{H2O} using the average value of available experimental self-broadening coefficients \citep{Ptashnik2016}.

% ???? CHECK SPACE BETWEEN A and B
%************************************************

\subsection{Modeling the Impact on Transmission/Emission Spectra of Transiting Exoplanets}

To assess the signifigance of the type-of-broadener assumption, we use the CHIMERA \citep{Line2013,Line2014,Stevenson2014a, Kreidberg2015a,LineParmentier2016,Kreidberg2018} code with our newly generated ACS (converted to $\lambda/\Delta \lambda$=100 correlated-K coefficients \citep{Amundsen2016}) to model transit/eclipse spectra of a representative sub-Neptune like planet (GJ1214b \citep{Harpsoe2013}, T$_{eq}$=500--900K). We first generate forward model spectra using both sets of ACS (\ce{H2O}@[self] and \ce{H2O}@[\ce{H2}+\ce{He}]) given a fixed temperature-pressure profile (TP, \cite{Guillot2010} Eqs. 24, 49 )\footnote{With $\kappa_{th}=3\times 10^{-2}$ cm$^2$/g, $\gamma=0.1$, T$_{eq}$=500,700,900K, T$_{int}$=0K} and either 100\% \ce{H2O} or 500$\times$Solar metallicity under thermochemical equilibrium\footnote{NASA CEA2 \citep{gordon1994} with scaled \citep{Lodders2009} abundances.  We include as opacities in this scenario \ce{H2}/He broadened \ce{H2O}, \ce{CH4}, CO, \ce{CO2}, \ce{NH3}, \ce{H2S}, Na, K, HCN, \ce{C2H2}, \ce{TiO}, VO, \ce{PH3}, and \ce{H2} \ce{H2}/He CIA \citep{Freedman2014a} }. Second, we compute a self-consistent radiative equilibrium atmosphere\footnote{Zero internal heat flux, PHOENIX stellar model for GJ1214, and an equilibrium temperature of 550 K so as to keep temperatures at all layers within the valid cross-section temperature range of 400--1500K} \citep{Arcangeli2018,Mansfield2018,Kreidberg2018} to determine the impact of water broadening on the vertical energy balance and, in turn, on the observed spectra.  We discuss our findings in the next section.
%%%%%%%%%%%%%%%%%%%%%%%%%%%%%%%%%%%%%%%%%%%%%%%%%%%%%%%%%%%%%%%%%%%%%%%%%%%%%%%%%%%%%%%%%
\section{Results}\label{Results}

%FIGURE 1******************************************
\begin{figure}
 \includegraphics[width=0.47\textwidth] {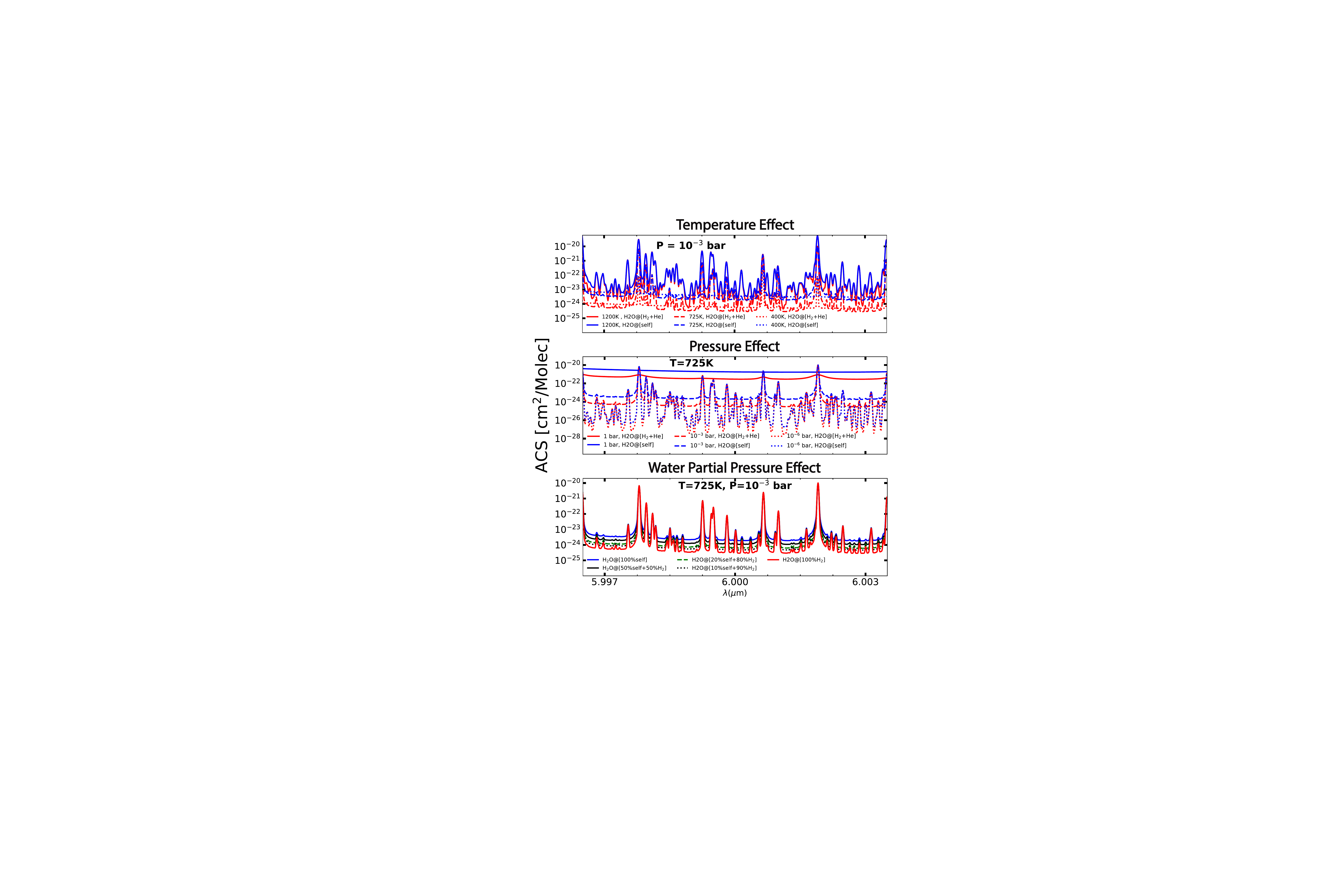} 
	  \caption{Illustration of the impact of @[self] (blue) versus @[\ce{H2}+He] (red) on the absorption cross sections near 6$\mu$m.  The top panel shows the influence of temperature on the broadening difference at a fixed represantative pressure of 1 mbar. At 1200K (1 mbar) the lines are purely Doppler broadened resulting in little effect.  The middle panel shows the influence of pressure at a fixed temperature. The Doppler cores are negligible by 1 bar.  The bottom panel shows the impact of the relative weighting of self versus \ce{H2} broadening  (e.g., composition dependence) at a fixed temperature and pressure.  Cross-section differences are largest in the pressure-broadened line wings, with pure @[self] typically 1 order of magnitude larger.  A factor of 5 in broadening difference occurs by the time the relative abundance of water reaches $\sim$30\%.  In general, @[self] broadening becomes more important at higher pressures, cooler temperatures, and longer wavelengths due to the increased prominence of pressure broadening over Doppler broadening.   }\label{fig:fig1} 
\end{figure}
%**************************************************

%FIGURE 2******************************************
\begin{figure*}[h!tb]
\centering
 \includegraphics[width=18.35cm,keepaspectratio] 
 		{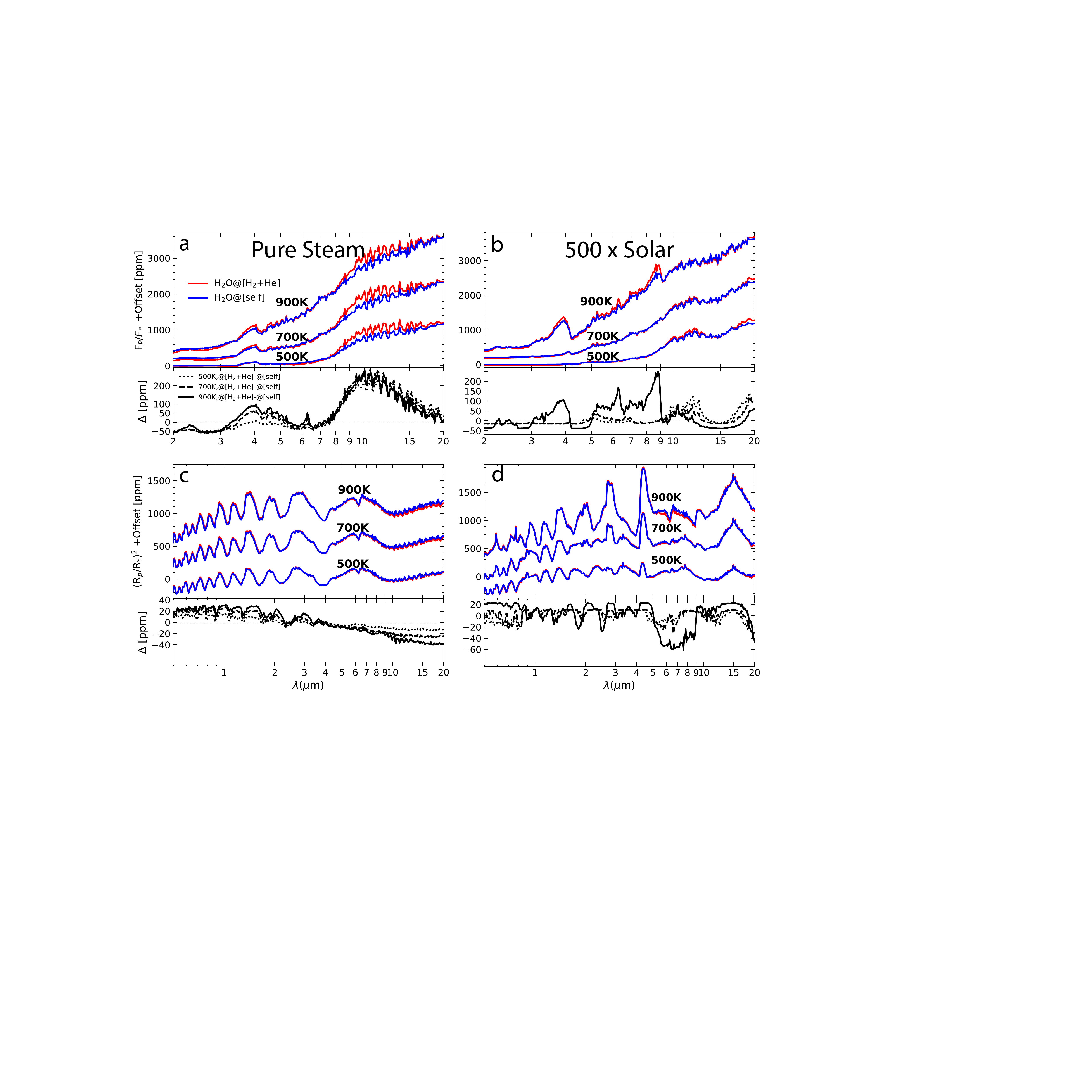} 
	  \caption{Effect of water self-broadening (@[self], blue) compared to the standard \ce{H2}/He broadening (@[\ce{H2}+\ce{He}], red) on pure steam (left column: a,c) and high metallicity (500$\times$ solar-right column: b,d) atmospheres with equilibrium temperatures of 500, 700, and 900K.  The top row (a,b) compares emission spectra and the bottom row shows relative transmission spectrum differences (c,d).  The bottom panel in each shows the relative spectral difference ($\Delta$).  Differences range anywhere from a few 10s to a few 100s of ppm and show a strong wavelength dependence. }\label{fig:fig2} 
\end{figure*}
%**************************************************

\subsection{Impact on Cross Sections}
Figure \ref{fig:fig1} illustrates the effect of temperature, pressure, and water abundance on the difference between @[self] and @[\ce{H2}+\ce{He}] broadened ACS near 6 $\micron$. The top panel shows how broadening changes with temperature at a fixed pressure of 1 mbar.  Differences are largest for cooler temperatures where pressure broadening becomes more important.  The middle panel illustrates the impact of variable pressure at a fixed temperature (725K). Even at low pressures (1 $\mu$bar) pressure broadening differences are still present in the line wings.  The bottom panel shows the effect of varying water abundance on the combined @[self]+@[\ce{H2}] broadening at a fixed temperature and pressure (725K, 1 mbar).  With pure self-broadening, differences in the line wings can approach an order of magnitude.  For a $\sim$30\% mole fraction of water, the ACS is about 3--5$\times$ greater than pure hydrogen broadening. While not shown, these differences become larger at longer wavelengths and smaller at shorter wavelengths due to the relative importance of Doppler-to-pressure broadening. 

\subsection{Direct Impact on Transmission/Emission Spectra}
More practically, Figure \ref{fig:fig2} summarizes the key impact of @[\ce{H2}+\ce{He}] versus @[self] broadening on the emission (top row) and transmission (bottom row) spectra of a typical sub-Neptune under the assumption of a pure steam atmosphere (left column) and a 500$\times$Solar metallicity\footnote{While the water mixing ratio is only $\sim$10--20\% for these conditions, we still use the pure @[self]-broadened water ACS as it is still a more accurate approximation than pure @[\ce{H2}+\ce{He}] broadening} scenario (right column).  Overall, we find that the differences are quite large, 10s to 100s of ppm, well within the detectable range of both HST \citep{Kreidberg2014a}, and certainly the James Webb Space Telescope (JWST, e.g., \cite{Greene2016,Bean2018}), especially for the anticipated windfall of such planets around bright stars \citep{Sullivan2015}.  

%% ALAN: There are hotter layers? What does the temperature in the figure correspond to?
%% ??? probably its just me but I don't understand this: 
In the all-steam atmospheres, emission differences (Figure \ref{fig:fig2}a) are largest in the window regions ($\sim 4\mu$m, $\sim 10\mu$m ). The increased flux for the @[\ce{H2}+\ce{He}] broadened ACS is because of the lower opacity, permiting flux from deeper, hotter layers to emerge (for a fixed TP).  The increased opacity due to the @[self] broadening obscures the deeper/hotter layers, resulting in lowered fluxes at those wavelengths.  These differences are, of course, strongly dependent upon the temperature structure within in the atmosphere.  As these spectra assume a fixed TP there is a difference in net radiated flux, which will most certainly have an influence on the radiative balance and thermal structure in the atmosphere, as discussed in \S \ref{sec:SelfCons}.

Transmission spectra tell a similar, albeit less dramatic story with relative differences of $\sim$60 ppm across shown wavelength range. The ``linear-like'' slope in the differences ($\Delta$) with wavelength is due to the frequency dependence of Doppler-to-Pressure broadening.  

The effects at high metallicity (500$\times$solar, Figure \ref{fig:fig2}, right column) are less extreme (10 of ppm) due to the reduced abundance of \ce{H2O} (10 -- 20\%) and the significant abundances of additional opacity sources (mainly \ce{CO2}, CO, \ce{CH4}, and \ce{H2}/He).   Furthermore, due to the reduced impact of \ce{H2O}@[self] broadening (Figure \ref{fig:fig1}), we expect an approximate (comparing 1 mbar line wings) reduction of 3--5$\times$ to $\sim <$ 10ppm in the transmission spectra. 

\subsection{Impact on Self-Consistent 1D Atmosphere}\label{sec:SelfCons}
\begin{figure*}[h!tb]
\centering
 \includegraphics[width=15cm,keepaspectratio] 
 		{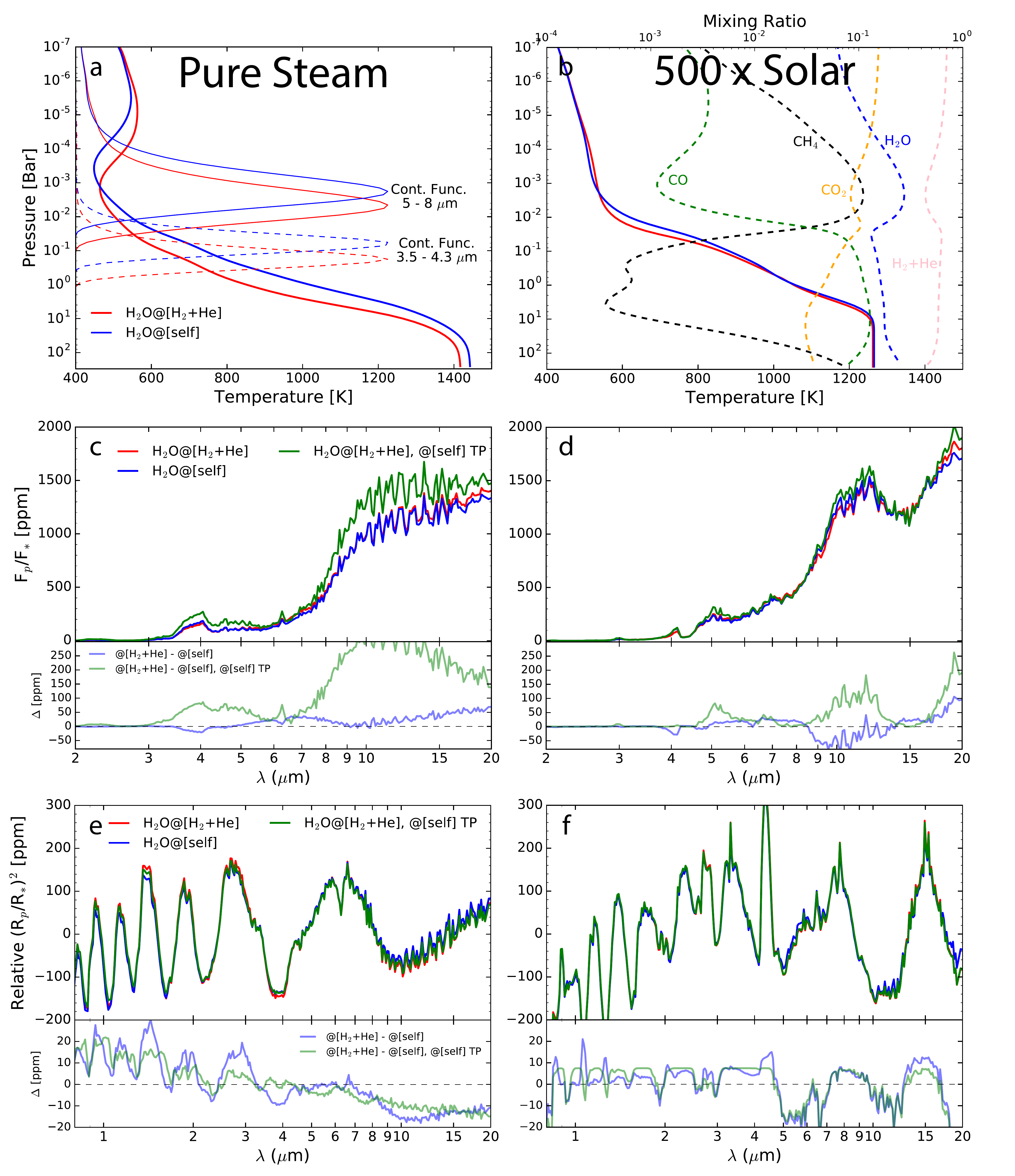} 
	  \caption{Comparison of the @[self] (blue) vs. @[\ce{H2}+He] (red) broadening in self-consistent 1D thermochemical-radiatve-equilibrium atmospheres for all steam (left column: a,c,e) and 500$\times$solar (right column: b,d,f) composition.  The top row (a,b) shows the derived radiative-equilibrium TP under each scenario.  Thermal emission contribution functions averaged over represenative bands (Cont. Func. 5--8, and 3.5--4.3 $\mu$m) for each broadening scenario are shown in (a). Subplot (b) shows the thermochemical equilibrium mixing ratios along the @[self] TP for select species. Temperature differences can be up to 175K (20\%) in the pure steam scenario and up to 70K in the 500$\times$Solar scenario.  The second row (c,d) shows the resultant secondary eclipse spectra and their differences below ($\Delta$). An additional emission spectrum (@[\ce{H2}+He], @[self] TP-green),  is shown in (c) and (d) assuming the same TP as the @[self] scenario in order to decouple the effects of the radiatively adjusted TP from the broadening differences. The last row (e,f) shows the resulting cloud free transmission spectra and relative differences.  An additional transmission spectrum (@[\ce{H2}+He], @[self] TP-green),  is shown in (e) and (f) assuming the same TP as the @[self] scenario in order to decouple the effects of the broadening and scale height change due to TP variation.  Spectral differences are on the order of 30-40 ppm in transmission but are much less in emission ($\sim$60 ppm) when compared to Figure \ref{fig:fig2}a,b due to the radiative adjustment of the TP.       } 
     %for steam integrated flux: (@H2/He-@[self]@[self], TP)/@H2/He=0.42, or eq Temp of 605 vs. 555 K
     %for 500x integrated flux: (@H2/He-@[self]@[self], TP)/@H2/He=0.09, or eq Temp of 567 vs. 555 K
      \label{fig:fig3}
\end{figure*}

Figure \ref{fig:fig3} shows the impact of self-broadening on the 1D radiative balance (and subsequent observational effects) of a $\sim$550K planet under the all steam and 500$\times$Solar scenarios.  The @[self] broadening results in $\sim$100-180K hotter temperatures below the $\sim$1 mbar level and $\sim$60Kcooler above for the all steam scenario (Figure \ref{fig:fig3}a).  More intuitively, the increased @[self] mean opacity ``shifts'' the averaged thermal ``$\tau$=1'' level to a $\sim 3\times$ lower pressure in the all steam scenario.  This shift is readily seen in the band averaged contribution functions (Figure \ref{fig:fig3}a). A similar, but lesser, effect is seen in the 500$\times$Solar metallicity scenario (up to $\sim$ 70 K) because the water abundance is lower by a factor of $\sim 5$ (Figure \ref{fig:fig3}b).  The radiative response of the TP to the integrated flux differences (up to 40\% for steam and 10\% for 500$\times$solar, green vs. red curves in Figure \ref{fig:fig3}c,d ) between the @[self] vs. @[\ce{H2}+He] acts to reduce the emission spectrum differences, however, to a still detectable 10s of ppm (Figure \ref{fig:fig3}c,d).

The transmission spectra (Figure \ref{fig:fig3}e,f) show comparable differences (30--40 ppm) to the 500 K scenario from Figure \ref{fig:fig2}c,d.  However, there are now two effects taking place that create the transmission differences. The first is the scale height effect due to the differences in the TP (@[\ce{H2}+He]-@[self], \ce{H2O} TP).  The second, as before, is the broadening differences.  Both effects contribute equally to the overall differences in the transmission spectra.
%% ???? ALAN- Composition change is another reason????
Despite the self-consistent adjustment of the TP, differences in both emission and transmission are still above detectable levels (10s of ppm)

%FIGURE 3******************************************
%% ALAN: also a long caption, The meaning of the green curves in c,d,e,f is not completely clear to me. This assumes constant TP as in Fig 2? The curves look a little bit different.
%**************************************************

%%%%%%%%%%%%%%%%%%%%%%%%%%%%%%%%%%%%%%%%%%%%%%%%%%%%%%%%%%%%%%%%%%%%%%%%%%%%%%%%%%%%%%%%%%%%%%%%%%%%%%%%%%
\section{Conclusions}\label{Conclusions}
%Random paragraph or something to insert into "Discussion" section, or maybe some of this can go in the intro as part of some "observational motivation".
%% ???? ALAN: perhaps another point to describe how part of the method is to test the use of incorrect cross sections
The aim of this work was to determine the observable impact of broadener composition on observed transiting planet spectra, with application to $<1000$K high-metallicity and all steam atmospheres, likely representative of the sub-Neptune/Super-Earth population of planets.  As a specific example, we focused on the difference between \ce{H2}/He broadening and self broadening on the water absorption cross-sections as water is typically the most prevalent species and absorber in planetary atmospheres.  From our analysis we arrive at the following key points:
\begin{itemize}
  \item Absorption cross section differences between water self and the standard assumed \ce{H2}/He broadening are up to an order of magnitude in the pressure broadened line wings (similar to \cite{Hedges2016}), and is noticeable over a range of applicable temperatures and pressures.
  \item The influence of self-broadening is composition dependent and non-linear, with $\sim$half of the difference achieved by water mole fractions of$\sim$30\% for a representative temperature and pressure.
  \item Transmission and emission spectra differences for representative sub-Neptune atmospheres range between a few 10s of ppm up to 100s of ppm, depending upon wavelength, temperature, and water abundance.  These differences are not negligible considering currently achieved HST precisions of $\sim$15 ppm and possible precisions as low as a few ppm for JWST.  Differences will vary depending upon additional parameters like temperature gradient (for emission), planet-to-star radius ratio, and scale height.  
  \item The assumption of water self-broadening (or lack thereof) can have a significant impact on the 1D vertical energy balance, with temperature differences of up to 180K in pure steam atmospheres (or a half-a-decade lower pressure shift in the emission levels) and 10s of K in high metallicity atmospheres.
\end{itemize}

This work is certainly not an exhaustive exploration of all possible broadening (Table \ref{tab: Table1}) or planetary atmosphere conditions. However, it serves to illustrate that the broadener assumption can have a non-negligible impact on the observables and continues to illustrate the importance and key role of laboratory data on planetary atmosphere modeling. \citep{Fortney2016}

\section{Acknowledgements}
EGN and MRL thank J. Lyons, A. Heays, R. Freedman, M. Marley, J. Fortney, P. Molli\`ere, and L. Pino for many useful discussions. We especially thank S. Yurchenko for invaluable assistance with the EXOCROSS code, and ASU Research Computing center for the kind support on computational side. MRL acknowledges summer support from the NASA Exoplanet Research Program award NNX17AB56G.  This work benefited from numerous conversations at the 2018 Exoplanet Summer Program in the Other Worlds Laboratory (OWL) at the University of California, Santa Cruz, a program funded by the Heising-Simons Foundation.

\bibliography{reference_7}

\end{document}